\documentclass[namedreferences]{solarphysics}

\usepackage[hyperref,optionalrh]{spr-sola-addons} 
\usepackage{graphicx}        
\usepackage{color}           
\usepackage{breakurl}        

\newcommand{\jgr}{    {\it J. Geophys. Res.}}
\newcommand{\mnras}{  {\it Mon. Not. Roy. Astron. Soc.}}

\newcommand{\solphys}{{\it Solar Phys.}}
 
\newcommand{\ssr}{    {\it Space Sci. Rev.}} 
\chardef\us=`\_

\begin{document}

\begin{article}
\begin{opening}

\title{Dark dots on the photosphere and counting of the sunspots index }

\author[addressref=aff1,email={tlatov@mail.ru}]{\inits{A.G.}\fnm{~Andrey~G.}~\lnm{~Tlatov}\orcid{123-456-7890}}

\runningauthor{A.G. Tlatov}
\runningtitle{Dark dots}

\begin{abstract}
On modern satellite observations of the Sun in the "continuum” with high spatial resolution, as well as on high-quality ground observations, a large number of small dark areas can be observed. These regions have no penumbra, have a contrast of up to $20\%$ and are similar to solar pores. The characteristic area of such structures is $0.3\div5\ \mu$hm or $0.5\div5$ Mm. The number of such points in one image can be several hundred. The nature of such formations remains unclear.

We have performed the selection of dark regions with a contrast of at least $3\%$ of the level of the quiet Sun on the SDO/HMI observational data in the continuum for 2010-2020. We have studied the properties of "dark points”, including the change with the cycle of activity, area distribution and contrast. We also compared such structures with the intensity of the magnetic field.   We found that the number of dark dots with an area of less than $5\ \mu$hm, in which the magnetic field is not significant and is less than $|B|<30$ G, is from 60 to $80\%$ of the total number of structures of this size. This means that these objects are not associated with magnetic activity. The existence of such structures can significantly affect the calculations of the sunspot index, since they can be mistaken as pores.
\end{abstract}
\end{opening}

\section{Introduction}
     \label{S-Introduction} 

Sunspots appear on the photosphere as regions with reduced temperature and radiation. The sunspot looks like a part of the photosphere with significantly less brightness than the surrounding areas.  Regular sunspots consist of a dark umbra surrounded by a lighter penumbra. During the transition from umbra to penumbra and photosphere, the intensity of radiation changes abruptly. Inside the penumbra, the intensity is approximately constant, inside the umbra, as a rule, it decreases towards the center. The umbra, on average, occupies $15 - 25\%$ of the spot area. The brightness of the umbra is $5 - 15\%$ of the brightness of the photosphere, but does not depend on the size of the umbra \citep{Vitinsky}.

  In addition to regular sunspots, there are pores and transitional spots \citep{Vitinsky, Tlatov14}.   {Small sunspots that do not have penumbra are called pores.} Its size ranges from $1”$ to $5”$ or $\sim 0.7 - 3.5\ Mm$, which corresponds to an area of $\sim 0.4 - 4\ \mu$hm \citep{Vitinsky}.  Pores are found both as individual formations and as part of groups of sunspots. The lifetime of the pores is several hours, but as part of a group, these can exist  up to several days. According to the observations of the Kislovodsk Mountain Astronomical Station of the Pulkovo   observatory (KMAS), 
  
  {The maximum value of the area distribution is $S_{por} \sim 3.6\ \mu $hm for pores and $S_{sp}\sim 120\ \mu$hm $(d\sim 30”)$ for regular sunspots \citep{Tlatov19}.} Transitional sunspots, as a rule, have an area in the range of $20 - 100\ \mu$hm and are located in between the pores and regular spots, since the penumbra of these is not fully formed    \citep{Tlatov14, Tlatov19}.
  
  The Sunspot Index $Ri$ counts sunspots and pores \citep{Clette14, Svalgaard16}. Currently, it is necessary to limit the size of small dark structures, {especially with the automatic detection method}. Earlier, for ground-based observations of the 19-th and early 20-th century, such a restriction was natural, since small aperture telescopes were used, and the observation conditions did not allow observing with a spatial resolution better than $2-4”$. However, recently, due to the beginning of observations on spacecraft, as well as ground-based observations on telescopes with a relatively large aperture and registration on CCD detectors with a short exposure time, it has become possible to register a structure of the order of one arc second or less. 
  
  In this work, we have performed the selection of dark structures in the photosphere according to the observations of the space observatory SDO/HMI for the period 2010-2021.  A large number of such structures are not associated with magnetic activity. Consequently, this may introduce errors in the calculation of the sunspot index.
  
\section{Data and Data Processing}
\label{S-aug}
\subsection{Differences in sunspot data}

\begin{figure}    
   \vspace{0.05\textwidth}     
   \centerline{
               \includegraphics[width=0.49\textwidth,clip=]{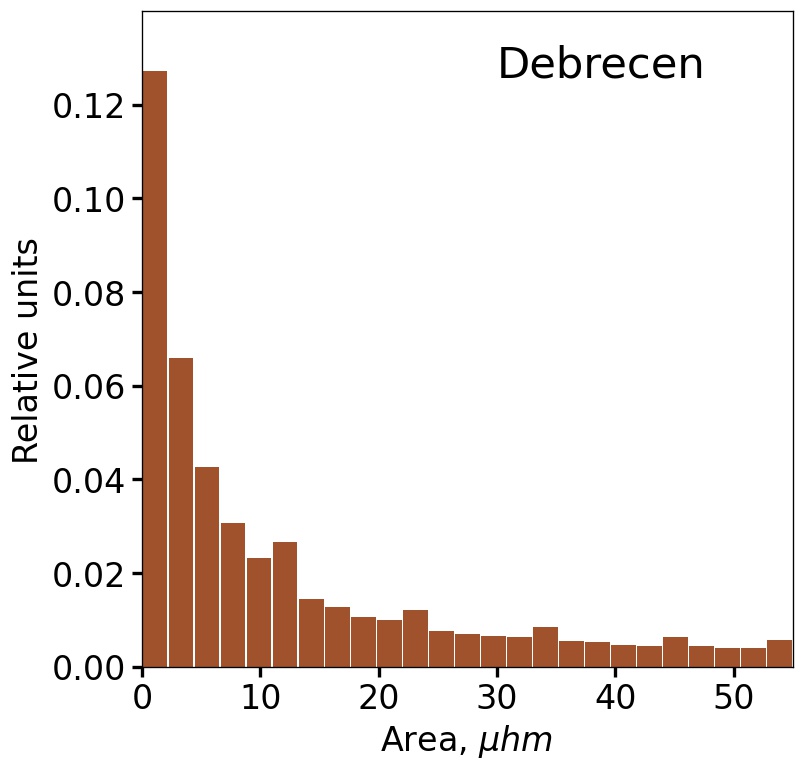}
               \includegraphics[width=0.505\textwidth,clip=]{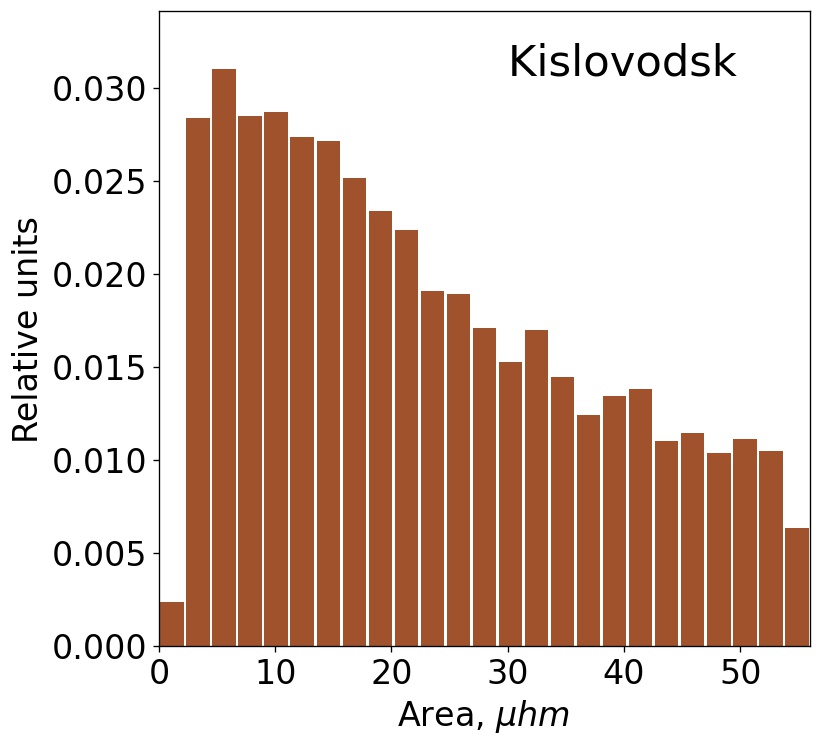}
              }
     \vspace{-0.51\textwidth}   
     \centerline{\Large \bf     
      \hspace{0.0 \textwidth}  \color{black}{(a)}
      \hspace{0.415\textwidth}  \color{black}{(b)}
         \hfill}
     \vspace{0.45\textwidth}    
     \caption{Distribution of the area of groups of sunspots for $S_{gr}<60\ \mu$hm according to the data of Debrecen (a) and Kislovodsk (b).}
 \label{Fig1}
   \end{figure}

Small-area groups and spots account for a significant part in the calculations of sunspot indices. Figure~\ref{Fig1} shows histograms of the distribution of the areas of sunspots  groups according to the data of the Debrecen \footnote[2]{\href{url}{http://fenyi.solarobs.csfk.mta.hu}}  and Kislovodsk \footnote[3]{\href{url}{http://www.solarstation.ru}} observatories for the period 1974-2018. 
{The maximum value of the distribution of the area} of Debrecen happen on groups with an area of $2 - 3\ \mu$hm. The percentage of groups with an area of $S<5\ \mu$hm in the total number is $\sim 25\%$. In the Kislovodsk data, the maximum area distribution is $8 - 10\ \mu$hm, and the proportion of groups with an area of $S<5\ \mu$hm is less than $3\%$.

The difference is significant.
{The observational data of the Debrecen Observatory were both its own observations and observations of Kislovodsk and other ground-based and satellite observatories.}
Both observatories use automatic sunspot recognition techniques \citep{Baranyi01, Tlatov14b}.  In Kislovodsk, spots with an area less than $2\ \mu$hm are not taken into account.   Therefore, the proportion of sunspots and small-area groups in Kislovodsk is significantly less than in the Debrecen data. But the question is, how justified is the choice of the lower threshold of the area $S_{min}> 2\ \mu$hm. 

We have performed SDO/HMI continuum image analysis.  The choice of such data from the space observatory provided high-quality images, without the influence of the atmosphere.
For the analysis, we used data from the SDO/HMI observatory of observations with an exposure of 45 seconds, at a time close to 5:00 UT.  The magnetic field data were selected for the same moment in order to most accurately combine measurements in the continuum with observations of magnetic fields.

For a comparative analysis of the characteristics of dark objects, it is necessary to know the typical contrast distributions for sunspots and pores. Observations of sunspots have been carried out at the Kislovodsk station since 1947. Since 2010, observations have been carried out on digital receivers.  Detection of sunspots, sunspot umbra and pores is carried out using a computer program in semi-automatic mode. In this case, the operator can change the conditions for selecting objects, which ensures the best accuracy. \citep{Tlatov14b}.  

\begin{figure}    
 \vspace{0.1\textwidth} 
   \centerline{
               \includegraphics[width=0.49\textwidth,clip=]{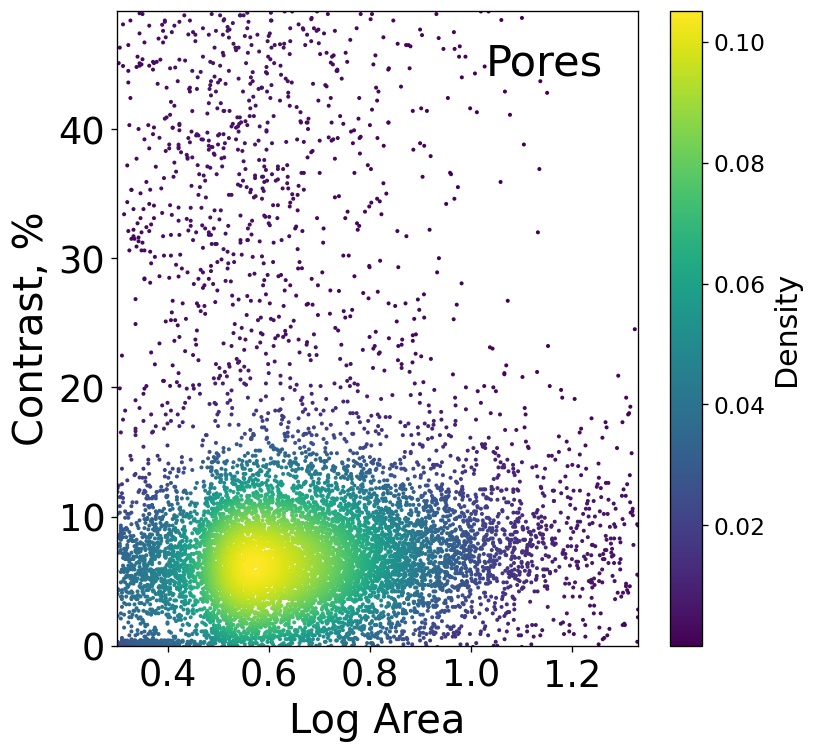}
               \includegraphics[width=0.4859\textwidth,clip=]{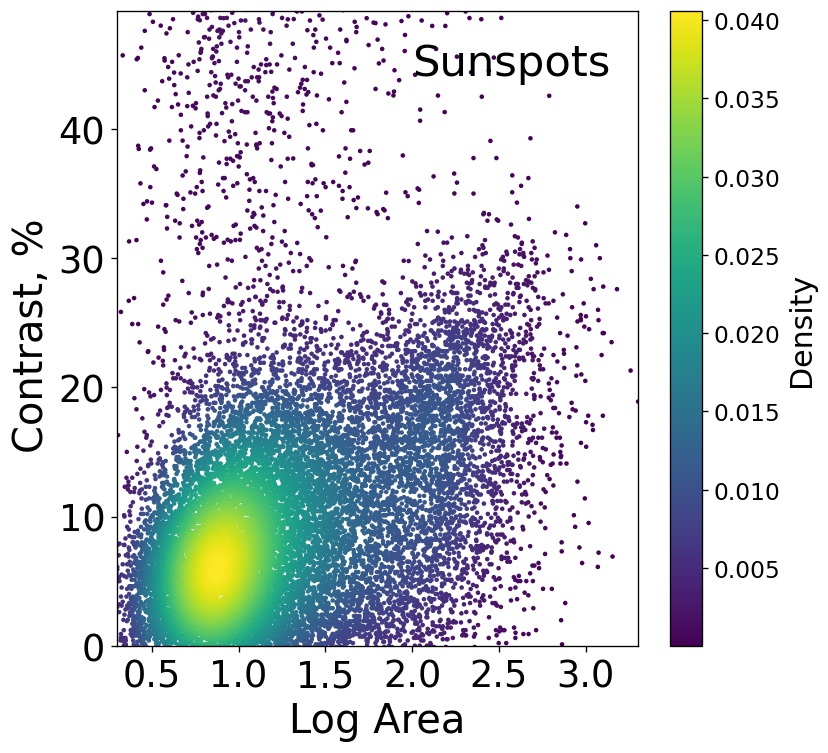}
              }
     \vspace{-0.53\textwidth}   
     \centerline{\Large \bf     
      \hspace{0.0 \textwidth}  \color{black}{(a)}
      \hspace{0.410\textwidth}  \color{black}{(b)}
         \hfill}
     \vspace{0.49\textwidth}    
     \caption{Contrast of pores (a) and spots (b) according to Kislovodsk data in the 24th cycle of activity depending on the logarithm of the area.}
 \label{Fig2}
   \end{figure}

{In \citep{Tlatov19}, we present} the characteristics of pores and spots according to KMAS observations. We used this series to determine typical contrast values.  Figure 2 shows the contrast of pores and sunspots (with penumbra) depending on the area. Contrast was calculated as the ratio of the minimum brightness in the object, relative to the brightness of the outer border. The brightness of the photosphere surrounding the object is taken as the value 1.  The pore contrast ranges from a few percent to $20 - 40\%$ (Figure~\ref{Fig2}a). The contrast of the spots can reach a much larger value (Figure~\ref{Fig2}b).
Thus, to detect pores and other dark objects, it is necessary to select objects with a contrast of at least several percent of the level of the undisturbed photosphere.  

To detect dark dots on SDO/HMI images, it is necessary to remove darkening to the edge and other possible intensity inhomogeneities on the disk.  The procedure for eliminating the darkening to the limb and determining the background intensity of $I_{bg}$ was as follows.  The disk of the Sun was divided into 12 segments at the polar angle. Inside each segment $k$, the intensity change function from the radius $R_{k}=f(i)$ was determined, where $i$ is the pixel number from the center of the disk.  In the center of the disk, for a circle of radius $R_{c}=0.3\cdot R$, the maximum intensity distribution $I_c$ was calculated.  For distances $r>R_{c}$, the background intensity was determined by interpolation $I_{bg}(r,\alpha)=I_{k}(r)\cdot a+b\cdot I_{k+1}(r)$, where the values of $a$ and $b$ were calculated depending on the angle $\alpha$ located between the centers of segments $k$ and $k+1$. For distances $R<R_{c}$, the background value was calculated as $I_{bg}(r,\alpha )=c\cdot I_{c}+d\cdot  I(r,\alpha)$.   The procedure was repeated twice. If at a distance $r$ there were regions of brightness exceeding $5\%$ of the average value, then at the second step they were excluded and the procedure was repeated. Then, the intensity {was recalculated in comparison to the intensity} at the center of the Solar’s disk $I_c$. To do this, the intensity of each pixel $(i,j)$ was calculated as $I^{\rm 2}_{i,j}=I_{c}-(I^{\rm o}_{i,j} -I^{\rm bg}_{i,j})$. This procedure makes it possible to eliminate darkening to the solar limb and, {to take} into account large-scale intensity inhomogeneities across the disk.

The procedure for selecting regions was as follows. A point with a minimum intensity was selected on the solar disk $(i,j)$. The intensity did not exceed of the average intensity $I_{i,j} <0.97\cdot I_c$. Next, we apply the growing procedure. In the vicinity of the point $(i,j)$, we found the intensity of the maximum of the distribution $I_{md}$ and the magnitude of the standard deviation $\sigma$. Next, we apply the growing procedure. To do this, we divide the intensity into intervals, with a step $\Delta I =0.5\cdot \sigma$. Further from the pixel $(i,j)$, we grow the area in the intensity range $I_{i,j} \div I_{i,j}+ \Delta I$. We attach bordering pixels from the selected range to the already selected pixels. Then we increase the upper limit of the intensity range by $\Delta I$ and repeat the growing procedure. To speed up the growing procedure, not all pixels are searched, but only those bordering on the selected ones that are not yet attached to the group. At each step k of the intensity range $I_{k} = I_{i, j} + k\cdot \Delta I$, we calculate the growth rate of the region by the number of newly attached points, compared with the number of points in the previous step. With an increase in the number of points $\Delta N = (N_{k}-N_{k-1}) / (N_{k-1} + N_{min})> 0.8$, the growing procedure stops. Here $N_{min}$ is the minimum number of pixels peculiar to the sunspot, for our case this size was determined from the minimum area of $S_{min}\sim 2\ \mu$hm.
Pixels attached at the last step $k: \Delta N>0.8$ are discarded. The pixels attached at steps $1\div k-1$ remain in the selected region. This procedure allows us to track the intensity of the boundary of the region $I_{rg}$. Since the rapid growth of the region at step $k$ means a transition from intensity from the local region to background values. Moreover, this $I_{rg}$ value can vary for different regions of the Sun's disk. This means that instead of some fixed threshold intensity level for the entire disk, we select a local threshold level specific to this part of the disk. In the future, we mark all the selected points of this region as already processed and exclude them from the further search procedure.

\begin{figure}    
   \centerline{\hspace*{0.015\textwidth}
               \includegraphics[width=0.5\textwidth,clip=]{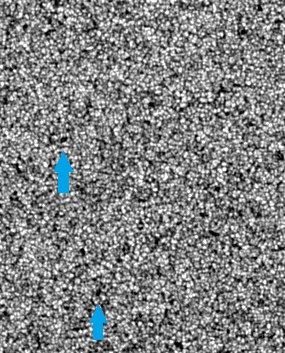}
               \hspace*{0.01\textwidth}
               \includegraphics[width=0.5\textwidth,clip=]{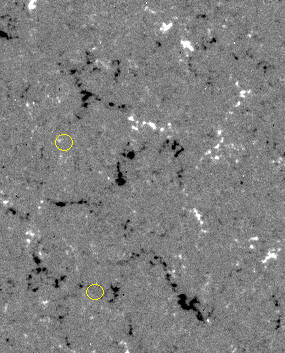}
              }
     \vspace{-0.60\textwidth}   
     \centerline{\Large \bf     
      \hspace{0.0 \textwidth}  \color{white}{(a)}
      \hspace{0.42\textwidth}  \color{white}{(b)}
         \hfill}
     \vspace{0.58\textwidth}    
     \caption{Image of a part of the solar disk in intensity (a) and magnetic field (b) for SDO/HMI 2010.07.04 5:00 UT. The arrows indicate dark areas for which there is no increase in the intensity of the magnetic field.   }
 \label{Fig3}
   \end{figure}

\begin{figure}    
  \includegraphics[width=1.0\textwidth,clip=]{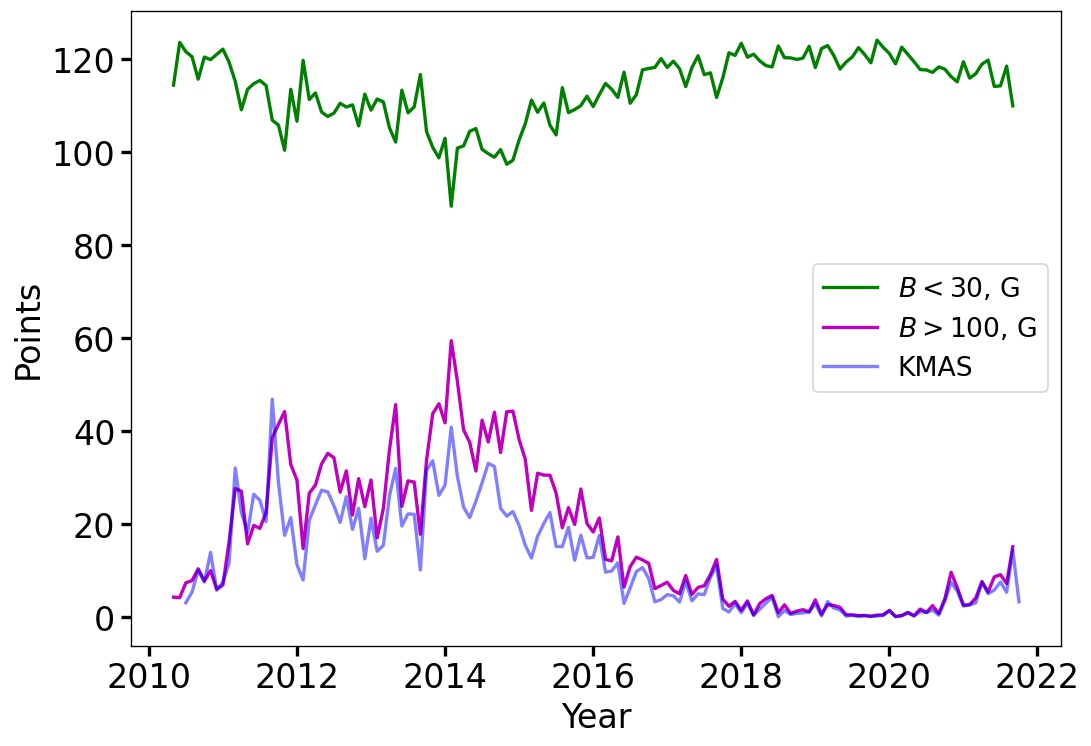}
   \vspace{0.1\textwidth}
   \includegraphics[width=1.0\textwidth,clip=]{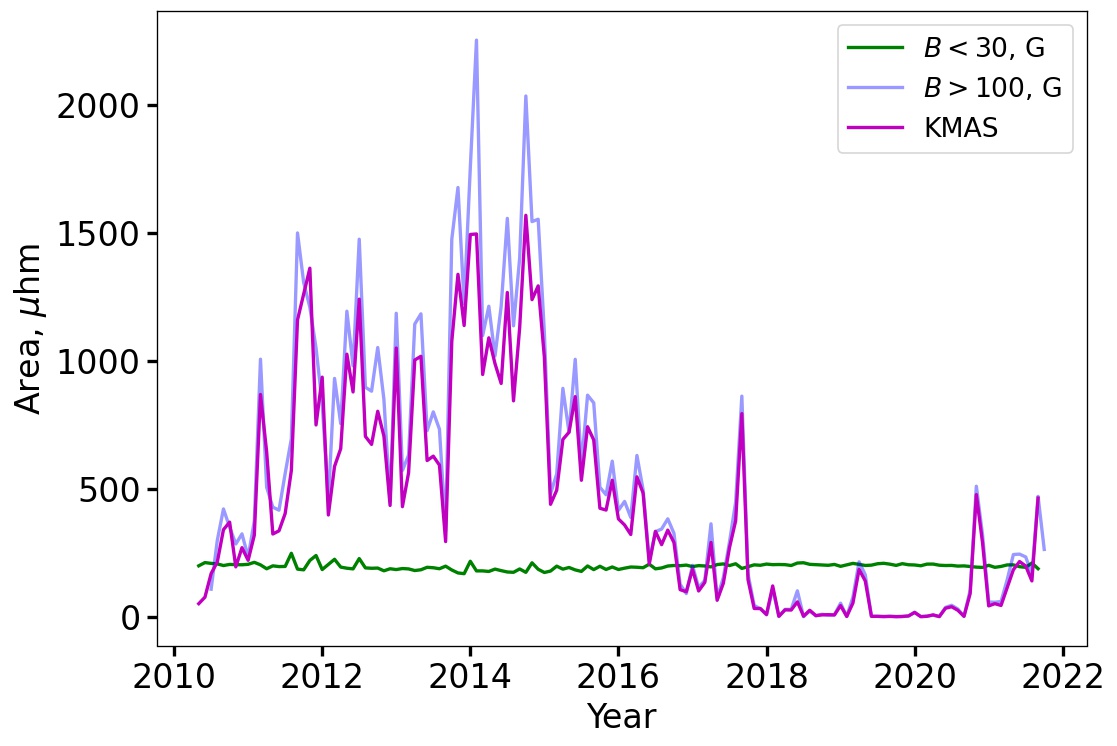}

     \centerline{\Large \bf     
         \hfill}
     \vspace{-0.1\textwidth}    
     \caption{Monthly average values of the selected dark structures of their number (top) and area (bottom). 
      {The results depend on} the size and intensity of the magnetic field.    }
 \label{Fig4}
   \end{figure}     

After that, we again look for points of minimum intensity and repeat the growing procedure. Sometimes two grown areas may touch borders. In this case, we combine them into one region.

Various parameters were determined for each region, including coordinates, area, average contrast, and other cluster parameters. The magnetic field was also measured. 

Figure~\ref{Fig3} shows images of a part of the solar disk in intensity and magnetic field for SDO/HMI 2010.07.04 5:00UT. In the figure, we have indicated dark regions for which there is no increase in the intensity of the magnetic field. There can be quite a lot of these regions on the disk.

\subsection{Data processing results}

Figure~\ref{Fig4} shows the average monthly values of the number and area of selected dark regions.     The area of dark regions is maximal in the epoch of maximum activity, and is close to the area of sunspots measured in Kislovodsk. We have identified structures in which the average magnetic field $B_{av}$ in absolute value is less than 30 G and more than 100 G. {These threshold values were chosen based on the magnitude of the average magnetic fields in sunspots and pores \citep{Tlatov14}}. If the total number of dark areas with a significant magnetic field is $40 - 60$ in the epoch of the maximum, then the number of areas in which the magnetic field is small is $100 - 120$ and varies slightly with the phase of activity. More precisely, their number is in antiphase with the number of regions having a significant magnetic field. This is probably due to the fact that at the maximum of activity, the area occupied by active regions in which magnetic fields are significantly growing.

The area of dark regions detected for SDO/HMI data by the fully automatic method  is close to the area of sunspots measured in Kislovodsk. The differences in the areas in the maxima of 2013 and 2014 are associated with omissions in ground observations and better accounting of the areas of large sunspots near the limb for satellite observations.

\begin{figure}[htp]    

{\Large \bf a)}
{
 \centering{}%
 \begin{minipage}[c][1\width]{0.9\textwidth}%
    \includegraphics[clip,width=1\textwidth]{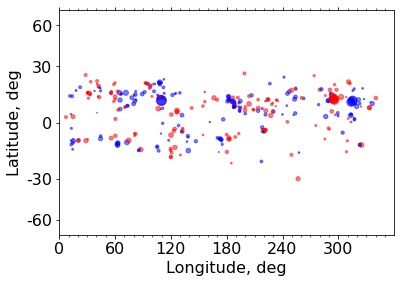}%
 \end{minipage}
}
 \vspace{-0.25\textwidth}
 
{\Large \bf b)}
{
\centering{}%
  \begin{minipage}[c][1\width]{0.93\textwidth}%
    \includegraphics[clip,width=1\textwidth]{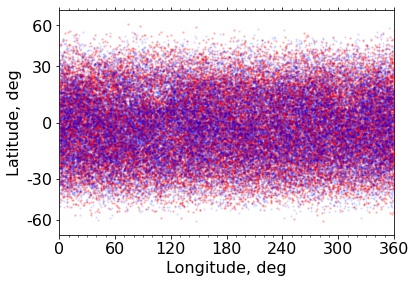}
  \end{minipage}
}
 
\vspace{-0.15\textwidth}
\caption{Distribution of dark areas on the summary synoptic map for 2017. a) for elements in which the magnetic field $|B_{av}|>100$ G and for b) for elements in which the magnetic field $|B_{av}|<30$ G. Red color corresponds to regions with a positive magnetic field sign, blue  with a negative one.     }
 \label{Fig5}
\end{figure}     

Figure~\ref{Fig5} shows summary synoptic maps of dark elements for elements with strong and weak magnetic fields for 2017. The distribution with strong magnetic fields corresponds to the localization of active regions.  Dark regions with weak magnetic fields evenly fill all longitudes and reach a latitude of $\sim 60^o$.

In our analysis, the contrast of detected dark areas exceeded $3\%$.  Figure 6 shows the dependences of contrast and maximum values of the magnetic field inside the region $B_{mx}$, separately near the active regions (AR) and outside them. To separate the regions, we considered that if the dark region was located at a distance of less than 8 degrees in longitude and 5 degrees in latitude from the center of the sunspots  group measured in Kislovodsk, then this region is connected by AR.
{These sizes roughly correspond to the largest AR in the 24-th cycle of activity, so if the distance is more than these values, then the region is outside the AR.}

\begin{figure}    

\vspace{0.10\textwidth}   

  \centerline{\hspace*{0.015\textwidth}
               \includegraphics[width=0.5\textwidth,clip=]{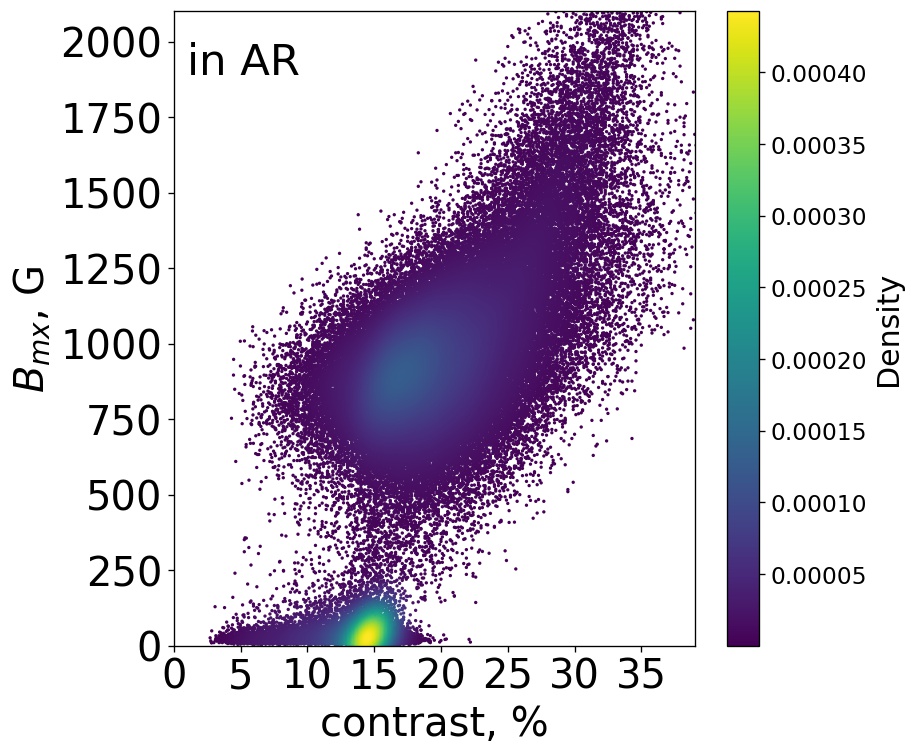}
               \hspace*{0.01\textwidth}
               \includegraphics[width=0.5\textwidth,clip=]{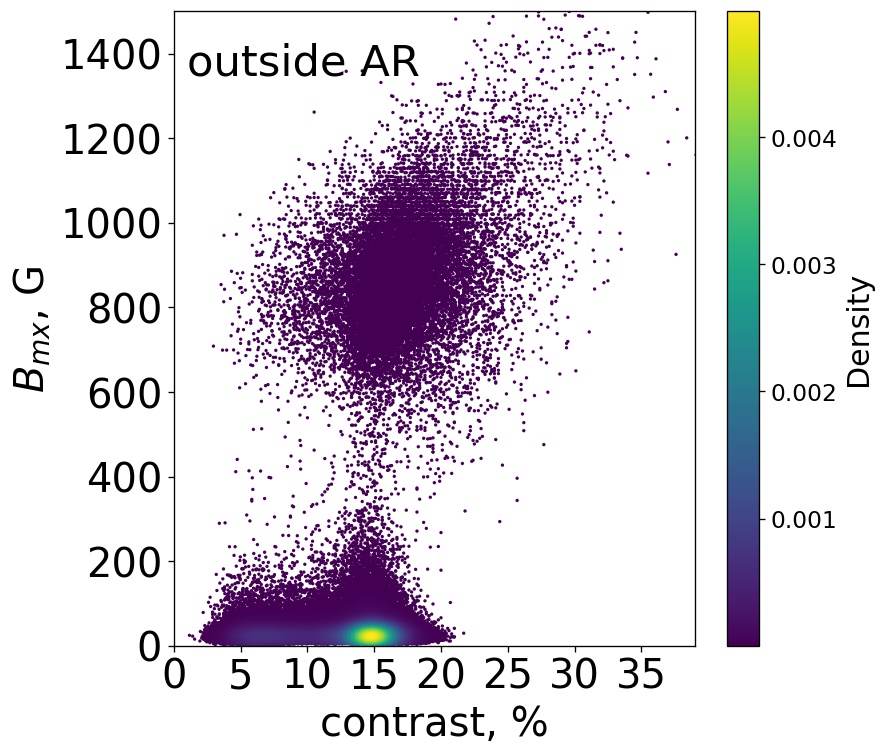}
              }
     \vspace{-0.52\textwidth}   
     \centerline{\Large \bf     
      \hspace{0.0 \textwidth}  \color{black}{(a)}
      \hspace{0.42\textwidth}  \color{black}{(b)}
         \hfill}
     \vspace{0.48\textwidth}    
     \caption{The relationship between the contrast of dark areas and the intensity of the magnetic field a) in AR; b) outside  AR. }
 \label{Fig6}
   \end{figure}

\begin{figure}    
 \vspace{0.09\textwidth} 
  \centerline{\hspace*{0.015\textwidth}
               \includegraphics[width=0.5\textwidth,clip=]{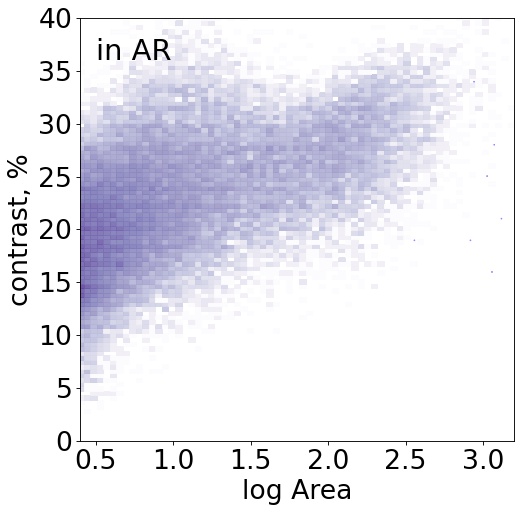}
               \hspace*{0.01\textwidth}
               \includegraphics[width=0.5\textwidth,clip=]{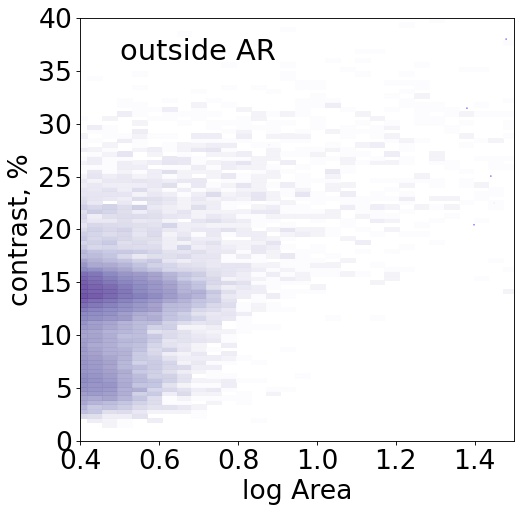}
              }
     \vspace{-0.54\textwidth}   
     \centerline{\Large \bf     
      \hspace{0.0 \textwidth}  \color{black}{(a)}
      \hspace{0.45\textwidth}  \color{black}{(b)}
         \hfill}
     \vspace{0.51\textwidth}    
     \caption{The relationship between the contrast of dark regions and the area. a) in AR; b) outside  AR.    }
 \label{Fig7}
   \end{figure}

Figure~\ref{Fig6} shows that there are two ranges of values for the magnetic field in the contrast range of $3 - 20\%$. The regions in which the magnetic field reaches $300 - 500$ G are obviously related to the magnetic fields of AR. But there are also quite a large number of dark regions in which the contrast is significant, and the magnetic field is not significant.  For contrast values of $\sim15\%$, there is a local maximum of distributions of the number of regions. With the increase in the area of dark regions that are located near AR, the contrast increases. Outside of AR, the contrast is weakly dependent on the area (Figure~\ref{Fig7}).

\begin{figure}    
  \centerline{
  \includegraphics[width=0.96\textwidth,clip=]{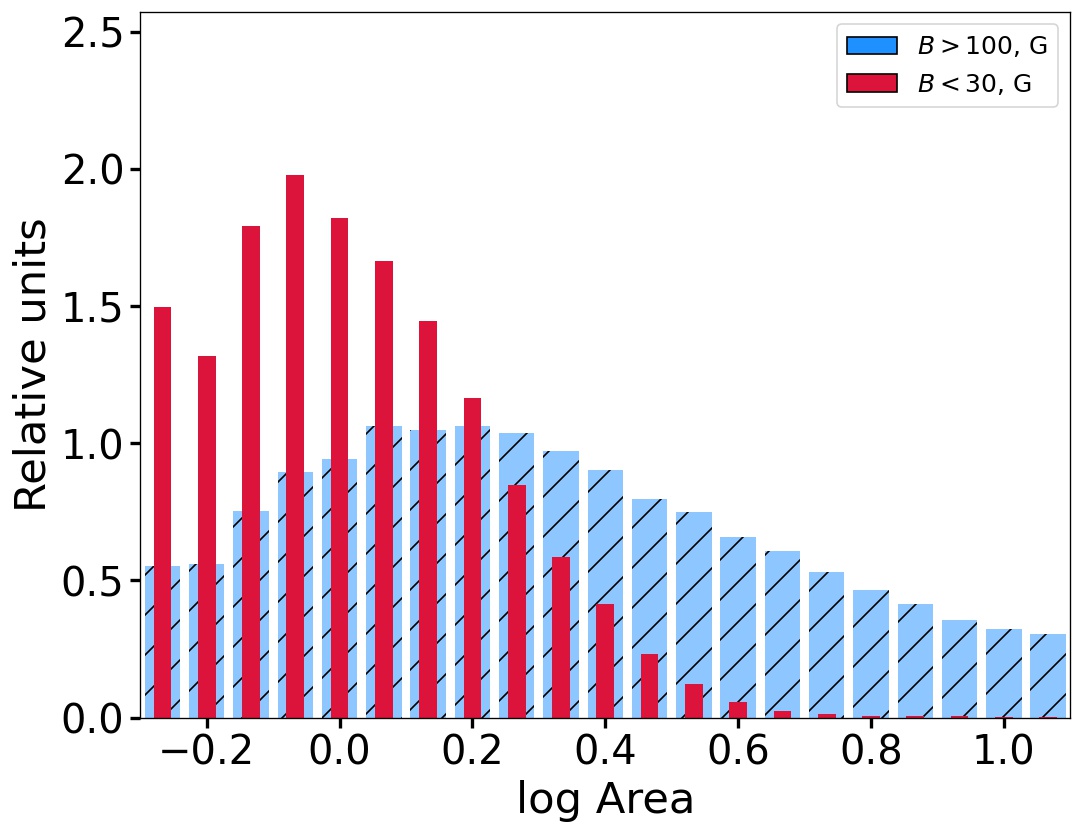}
  }
     \caption{Distributions of the relative number of dark regions from the logarithm of the area for regions with strong and weak magnetic fields.  }
 \label{Fig8}
   \end{figure}

\begin{figure}    
 \centerline{
  \includegraphics[width=0.96\textwidth,clip=]{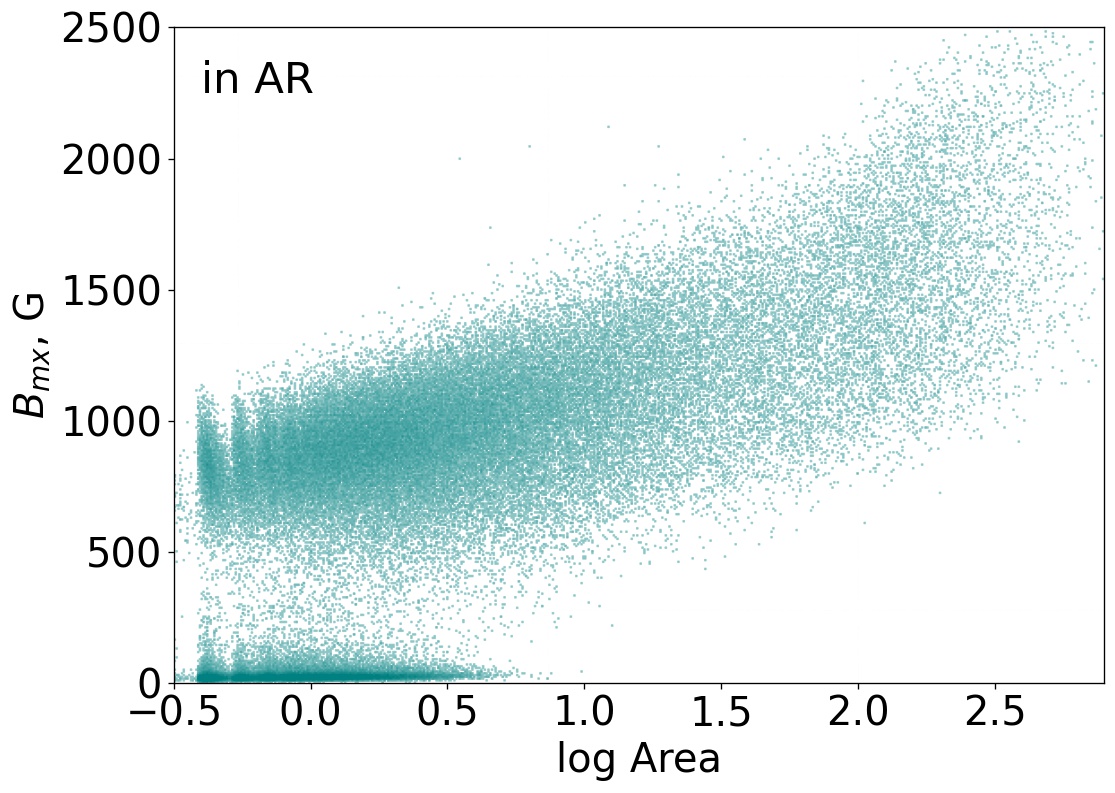}
 } 
     \caption{The relationship between the intensity of the magnetic field $B_{mx}$ and the logarithm of the area of dark regions. }
 \label{Fig9}
   \end{figure}

\begin{figure}    
 \centerline{
 \includegraphics[width=0.9\textwidth,clip=]{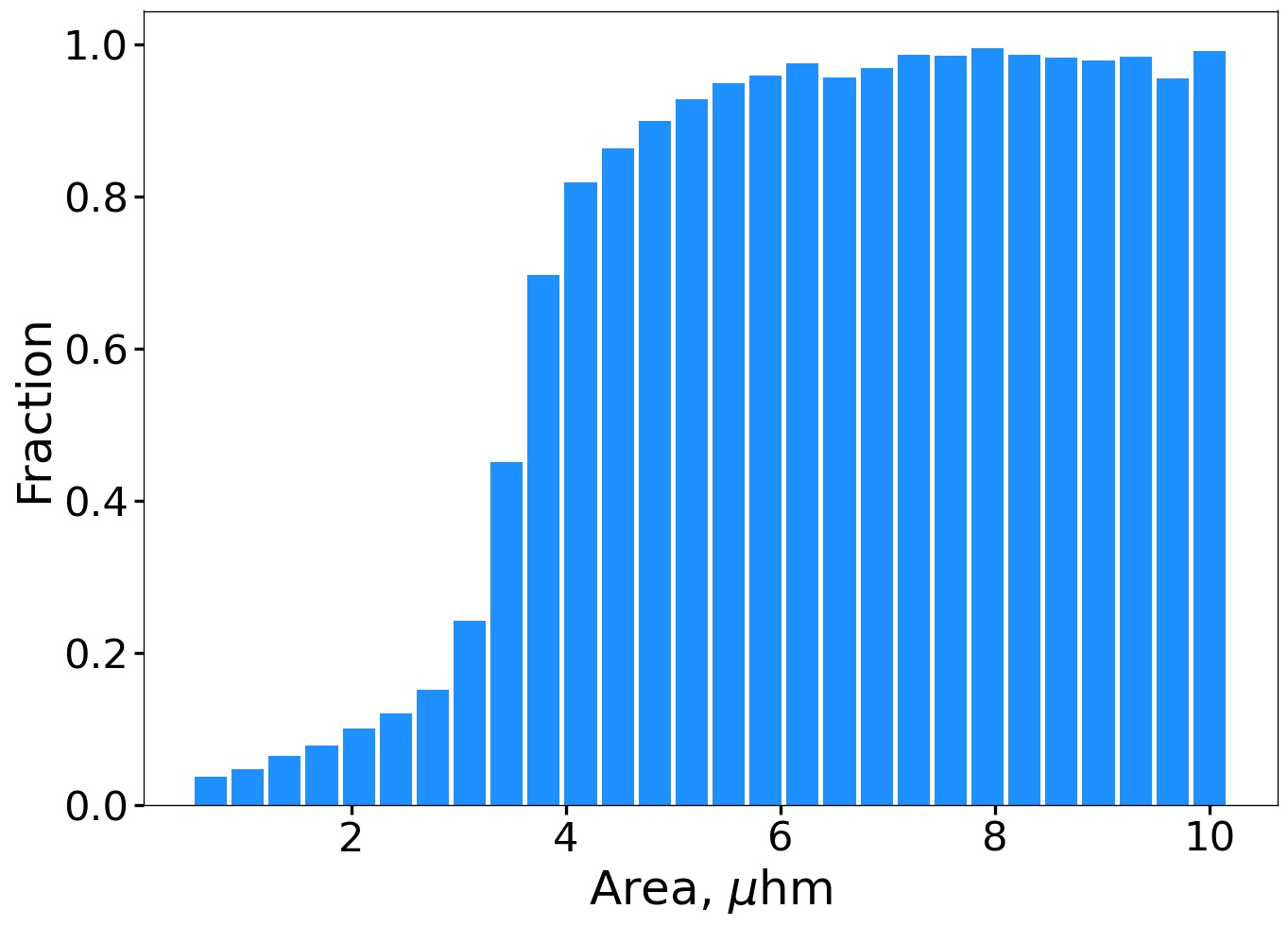}
 }
     \caption{The proportion of dark regions with a magnetic field $B_{av} >100$ G in the total amount, depending on the area in $\mu$hm. }
 \label{Fig10}
   \end{figure}     

{The distribution of dark region from the logarithm of the area  is shown in Figure~\ref{Fig8}. The maximum value of distribution of areas}
 with a strong magnetic field is $\sim 1.2 - 1.5\ \mu$hm. For regions with a weak magnetic field, the maximum distribution is $\sim 1\ \mu$hm.
 
   The distribution of the magnitude of the $B_{mx}$ magnetic field from the logarithm of the area is shown in Figure~\ref{Fig9}. For areas in which $B_{mx} > 500$ G, there is an increase in the intensity of the magnetic field with increasing area. This corresponds to the regularities for the magnetic fields of regular sunspots \citep{Tlatov14}. But there are also regions for which the magnetic field is small. For these, there is no change in the magnetic field with increasing area.
   
\section{Discussion}

We performed detection of dark regions on the photosphere based on observations in the SDO/HMI continuum. We used an automatic method of selecting such structures (Section 2.2). The use of SDO/HMI data made it possible to combine observations in "white" light with measurements of magnetic fields. This made it possible to establish that some of the dark regions do not have a significant magnetic field characteristic of sunspots and pores \citep{Tlatov14}.

The area of the selected dark regions according to SDO/HMI data, in which the magnetic field has large values, turned out to be close to the area of the spots detected at the KMAS observatory (Figure~\ref{Fig4}). This indicates the validity of the applied algorithm for spot detection.

Our analysis showed that the selection of dark regions on the photosphere in the continuum can lead to an erroneous interpretation of the magnetic activity of the Sun. Indeed, a large proportion of such small-sized regions are not associated with a magnetic field. Figure 10 shows for dark regions having a significant magnetic field $B>100$ G to the total number of detected regions. For regions with an area of less than $\sim4\div5\ \mu$hm, the proportion of regions with an insignificant magnetic field may be significant. Thus, for regions with an area of less than 3 $\mu$hm, the number of regions not associated with strong magnetic fields is more than $80\%$ (Figure~\ref{Fig10}).

In the article \citep{Tlatov19}, according to KMAS data, the characteristic pore area was determined in $S_{por}\sim 4\ \mu$hm. Our analysis showed that about $80\%$  {of dark regions of this area} have a significant magnetic field. However, according to the data of the Debrecen Observatory \citep{Baranyi01}, the maximum
value of the distribution of the area of groups of sunspots corresponds to 2 $\mu$hm (see Figure~\ref{Fig1}). But according to the analysis presented here, more than $80\%$ of such spots are not associated with a magnetic field. This is probably why the total number of spot groups in Debrecen is more than 2 times the number of spot groups in Kislovodsk, although the areas of sunspot groups are approximately equal \citep{Baranyi18}.
  
Thus, caution should be exercised when assigning small dark regions to the manifestation of solar activity. It may be necessary to limit the pore size in calculating the sunspot index. For single pores, this size should be limited to $S_{min}\sim 4\ \mu$hm (see Fig. 10). Figures 6 and 9 show that small regions without a magnetic field can exist not only far away, but also near AR. The area of such areas is usually up to $S_{min}\sim 2\div3\ \mu$hm. Therefore, less than this value, they should not be taken into account in the calculation of the sunspot index.  
 
Another method is the detection of pores using observations of magnetic fields, as is done in this work.  But this method significantly limits the use of simple photoheliograph telescopes. Another verification method may be to take into account small pores, only near already existing groups of sunspots or faculae visible in white light.
{We can also take into account the lifetime and latitude for sampling sunspots and pores.}

\section{Conclusion}

We performed an analysis of dark elements {present} in the continuum according to SDO/HMI observations and compared these elements with the magnetic field measured at the same time. It appeared that on the  images of the Sun's disk there are quite a lot ($\sim100$) elements with an area of $0.3 - 5\ \mu$hm (or a diameter of $0.5 - 5$ Mm) with a contrast of $\sim 3 - 20\%$, having weak magnetic fields and, apparently, not related to magnetic activity. These regions, which can be called "black dots”, can be mistaken for solar pores in which the intensity of magnetic fields exceeds several hundred Gauss \citep{Tlatov14}. The number of black dots decreases at the maximum of activity and increases at the epoch of the minimum of activity. The black dots are evenly distributed along the longitude and are located in the mid-latitude zone up to latitudes $50\div60^o$.

The existence of black dots should be taken into account when calculating of  the sunspots index, since they can be falsely mistaken for solar pores. This is especially important when analyzing images with good spatial resolution, such as satellite observations and observations at high-altitude observatories. The counting the sunspot index on such images, especially with the use of automatic processing methods, can significantly affect the stability of the sunspot index series.

{The rules should be adapted in case of
automatic segmentation of sunspots on high-resolution images.}
Probably, single pores should not be considered if their size is less than $4\ \mu$hm. When observing near the limb, small pores, including clusters, should not be taken into account if they exist without accompanying faculae. {The counted pores should be limited to an area of $\sim 2\ \mu$hm.}

\end{article} 

\end{document}